\documentclass[twoside,showpacs,superscriptaddress,twocolumn,floatfix,reprint,a4paper]{revtex4}

\usepackage{color}

\usepackage{latexsym}
\usepackage{graphicx}
\usepackage{times}
\usepackage{bbold}

\newcommand{\subtxt}[1]{\mbox{\tiny #1}}
\newcommand{\qnula}{0}
\newcommand{\qjedna}{1}

\begin{document}

\title{Preparation of Knill-Laflamme-Milburn states using tunable controlled phase gate}

\author{Karel\ Lemr}

\address{RCPTM, Joint Laboratory of Optics of Palacky University and Institute of Physics of Academy of Sciences of the Czech Republic, Faculty of Science, Palacky University \\17. listopadu 12, 77146 Olomouc, Czech Republic}

\date{\today}

\begin{abstract}
A specific class of partially entangled states known as Knill-Laflamme-Milburn states (or KLM states) has been proved to be useful in relation to quantum information processing \cite{knill01}. Although the usage of such states is widely investigated, considerably less effort has been invested into experimentally accessible preparation schemes. This paper discusses the possibility to employ a tunable controlled phase gate to generate an arbitrary Knill-Laflamme-Milburn state. In the first part, the idea of using the controlled phase gate is explained on the case of two-qubit KLM states. Optimization of the proposed scheme is then discussed for the framework of linear optics. Subsequent generalization of the scheme to arbitrary $n$-qubit KLM state is derived in the second part of this paper.
\end{abstract}


\maketitle

\section{Introduction}
Important developments have been demonstrated in quantum information processing (QIP) in the past few decades \cite{alber01,bowmeester00,nielsen02}. Several outcomes of this scientific field such as quantum cryptography \cite{benett84,benett92,eckert91,gisin02} or random number generation \cite{inoue83,peres92,jennewein00,stefanov00,katso08} have already found their industrial applications. In other cases a lot of effort has yet to be invested into the research. Mainly the lack of some experimental tools (e.g. strong optical non-linearity \cite{turchette95}) prevents from developing efficient quantum devices. An important discovery has been achieved by Knill, Laflamme and Milburn \cite{knill01}, when they have derived that a specific class of partially entangled states (so called Knill-Laflamme-Milburn states, or simply KLM states) can be used to significantly improve the efficiency of quantum computing. They have proposed a nearly deterministic teleportation based protocol for quantum computation using the KLM states as ancillas. In this protocol the overall success probability of quantum computation goes asymptotically to unity with growing number of photons in the ancillary KLM state. Their work has been followed by several other related proposals and experiments \cite{franson02,okamoto07,grudka08}. Franson~\it et al.\,\normalfont \cite{franson02} have generalized the original KLM scheme so that the success probability of quantum computing scales better with growing number of photons, but at the expense of lower fidelity of the output states. Several schemes for preparation of KLM states have also already been proposed. The general preparation idea has been mentioned in the original KLM paper \cite{knill01} though there was no specific recipe. The first explicit scheme for preparation of the KLM states was proposed by Franson~\it et al.\,\normalfont and it uses non-deterministic controlled sign gates and single photon interference to generate arbitrary photon-number KLM states \cite{franson04}. Another scheme limited only to 2-photon KLM states, but not requiring any post-selection, was also proposed \cite{lemr08} and subsequently experimentally implemented \cite{lemr10}.

This paper investigates yet another approach for experimentally accessible preparation of KLM states using the controlled phase gate (c-phase gate). The advantage in using this gate is the fact that the c-phase gate is considered an important part of the QIP toolbox \cite{sleator95,barenco95}. The Franson~\it et al.\, \normalfont scheme also employs the controlled phase gates (or in their case controlled sign gates) but with constant phase shift set to $\pi$. In this paper a fully tunable controlled phase gate is considered and a scheme for it's usage as a resource for KLM state generation is developed. By this strategy the overall success probability of the KLM state preparation can be increased considerably for some KLM states as it is shown in this paper. The presented scheme is fully general and allows to prepare KLM states of arbitrary number of qubits. Also no previous entanglement between the input qubits is required as the entangling capability of the gate itself is sufficient. The fully tunable controlled phase gate capable of imposing any phase shift in the range from 0 to $\pi$ has already been both proposed theoretically \cite{kieling10} and implemented experimentally \cite{lemr11} on the platform of linear optics and thus can be considered experimentally accessible.

\section{Basic 2-qubit scheme}
Using the qubit representation, one can express the $n$-qubit KLM state in the form of
\begin{equation}
\label{eq:multiKLM}
|\psi\rangle_{\subtxt{KLM}} = \sum_{j=0}^n \alpha_j |\qjedna\rangle^j |\qnula\rangle^{n-j}.
\end{equation}
The original definition by Knill, Laflamme and Milburn sets $\alpha_j = \frac{1}{\sqrt{n+1}}$ for $j = 0,...,n$, but the subsequent research carried out by Franson~\it et al.\normalfont\, \cite{franson02} indicates, that additional benefits can be found in using general amplitudes $\alpha_j$. Their research revealed that one can increase the efficiency of teleporation based quantum computing for instance by choosing triangular shaped amplitudes $\alpha_j$ (that is $\alpha_0 = \alpha_n = 0$ and alpha linearly growing towards maximum at $\alpha_{n/2}$ and then decreasing). This improvement is obtained at the expense of lower fidelity of the output state. (For more details please consult \cite{franson02}).

\begin{figure}
\includegraphics[scale=1.1]{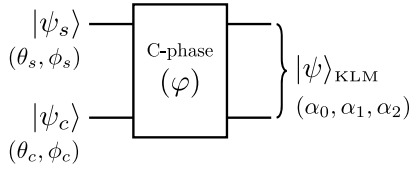}
\caption{Scheme of the proposed procedure for generation of two-qubit KLM states. The signal and control input qubit undergo a c-phase gate with tunable phase shift $\varphi$ yielding the two-qubit KLM state.}
\label{fig:scheme}
\end{figure}

In the first part of this paper let us consider the preparation of two-qubit KLM states (see figure \ref{fig:scheme}). The generalization to an arbitrary number of qubits would be presented later. Using the general definition for the KLM states (\ref{eq:multiKLM}) one can find that the two-qubit KLM states are in the form of 
\begin{equation}
\label{eq:klm_qubit}
|\psi\rangle_{\subtxt{2-QUBIT KLM}} = \alpha_0 |\qnula\qnula\rangle + \alpha_1 |\qjedna\qnula\rangle + \alpha_2 |\qjedna\qjedna\rangle,
\end{equation}
where $\alpha_j$ (for $j=0,1,2$) are arbitrary complex amplitudes following the normalization condition $\sum_{j=0}^2|\alpha_j|^2 = 1$. Having the target state well defined let us now inspect the properties of the c-phase gate.

The c-phase gate is a two-qubit quantum gate whose action in the gate's computational basis reads
\begin{eqnarray}
\label{eq:cphase}
|00\rangle &\rightarrow & |00\rangle \nonumber\\
|01\rangle &\rightarrow & |01\rangle \nonumber\\
|10\rangle &\rightarrow & |10\rangle \nonumber\\
|11\rangle &\rightarrow & \mbox{e}^{i\varphi} |11\rangle 
\end{eqnarray}
with numbers in brackets denoting first and second qubit state. General c-phase gate can be set to impose an arbitrary phase shift $\varphi$ to the two-qubit state $|\qjedna\qjedna\rangle$.

Any signal and control qubit can be expressed in terms of the gate's computational basis
\begin{equation}
\label{eq:cphase_input}
|\psi_{c,s}\rangle = \cos{\theta_{c,s}} |\qnula_{c,s}\rangle + \mbox{e}^{i\phi_{c,s}} \sin{\theta_{c,s}} |\qjedna_{c,s}\rangle,
\end{equation}
where indexes $c$ and $s$ denote the control and signal qubit. Please note that this state can always be prepared with high fidelity using only single qubit transformations (e.g. wave-plates in the case of photon polarization encoding). The separable input state $|\psi_c\psi_s\rangle$ is transformed by the gate yielding
\begin{eqnarray}
\label{eq:cphase_output}
|\psi\rangle_{\subtxt{OUT}} & = & \cos{\theta_c} \cos{\theta_s} |\qnula\qnula\rangle + \mbox{e}^{i\phi_{s}} \cos{\theta_c}\sin{\theta_s} |\qnula\qjedna\rangle + \nonumber\\
 & & + \mbox{e}^{i\phi_{c}} \sin{\theta_c}\cos{\theta_s} |\qjedna\qnula\rangle + \nonumber\\
 & & + \mbox{e}^{i(\phi_{c}+\phi_{s}+\varphi)} \sin{\theta_c}\sin{\theta_s} |\qjedna\qjedna\rangle.
\end{eqnarray}
Using the expression for signal qubit (\ref{eq:cphase_input}), the output state can be rewritten to the following form
\begin{equation}
\label{eq:cphase_output_alter}
|\psi\rangle_{\subtxt{OUT}} = \cos{\theta_c} |\qnula\psi_s\rangle + \mbox{e}^{i\phi_{c}} \sin{\theta_c} \left( \tau |\qjedna\psi_s\rangle + \epsilon |\qjedna\psi^\bot_s\rangle\right),
\end{equation}
where $|\psi^\bot_s\rangle$ is the orthogonal state to $|\psi_s\rangle$ so that $\langle\psi^\bot_s|\psi_s\rangle = 0$ and the parameters $\tau$ and $\epsilon$ are defined as
\begin{eqnarray}
\label{eq:tau_epsilon}
\tau & = & \langle\psi_s|\left(\cos{\theta_s}|\qnula\rangle + \mbox{e}^{i(\phi_{s}+\varphi)} \sin{\theta_s} |\qjedna\rangle \right) = \nonumber\\
& = & \cos^2{\theta_s} + \mbox{e}^{i\varphi} \sin^2{\theta_s}, \nonumber\\
\epsilon & = & \langle\psi^\bot_s|\left(\cos{\theta_s}|\qnula\rangle + \mbox{e}^{i(\phi_{s}+\varphi)} \sin{\theta_s} |\qjedna\rangle \right) = \nonumber\\
 & = & \mbox{e}^{i\phi_s} \sin{\theta_s} \cos{\theta_s} \left(1-\mbox{e}^{i\varphi}\right).
\end{eqnarray}
After performing the single qubit transformation
\begin{equation}
\label{eq:signal_transform}
|\psi_s\rangle \rightarrow |\qnula\rangle, |\psi^\bot_s\rangle \rightarrow |\qjedna\rangle
\end{equation}
in the signal mode, one can clearly recognize the two-qubit KLM state in the output state of the gate
\begin{equation}
\label{eq:cphase_output_klm}
|\psi\rangle_{\subtxt{OUT}} = \cos{\theta_c} |\qnula\qnula\rangle + \mbox{e}^{i\phi_{c}} \tau \sin{\theta_c} |\qjedna\qnula\rangle + \mbox{e}^{i\phi_{c}} \epsilon \sin{\theta_c} |\qjedna\qjedna\rangle,
\end{equation}
The remaining task is to map the complex amplitudes in (\ref{eq:cphase_output_klm}) to the original amplitudes $\alpha_j$ and to show that any two-qubit KLM state is achievable.

First let us consider the relative amplitude ratio and phase between $\alpha_0$ and $(\alpha_1 + \alpha_2)$. Any amplitude ratio can easily be set just by the choice of the $\theta_c$ parameter of the input control state
\begin{equation}
\label{eq:a_bg}
\frac{|\alpha_1|^2 + |\alpha_2|^2}{|\alpha_0|^2} = \tan^2\theta_c.
\end{equation}
As for the phase, the freedom in setting any value of $\phi_c$ assures that any phase shift between $\alpha_0$ on one side and $\alpha_1$ and $\alpha_2$ on other side is achievable.

The relation between $\alpha_1$ and $\alpha_2$ is also simple. For instance setting the phase shift $\varphi = \pi$ simplifies the amplitude ratio to
\begin{equation}
\label{eq:g_b}
\frac{|\alpha_2|}{|\alpha_1|} = \frac{|\epsilon|}{|\tau|} = \tan 2\theta_s
\end{equation}
and an arbitrary phase shift between $\alpha_1$ and $\alpha_2$ can be set by the choice of $\phi_s$. Please note that setting $\varphi = \pi$ allows to cover the whole class of KLM states. This fact will be used for the discussion in section 5. The equations (\ref{eq:a_bg} and \ref{eq:g_b}) manifest that any amplitude ratio between $\alpha_0$, $\alpha_1$ and $\alpha_2$ is achievable since $\tan$ goes from 0 to $\infty$.

\section{Success probability optimization}
One may conclude that the tunability of the gate in the phase shift $\varphi$ is a redundant feature. However this parameter can be used for optimization of the procedure. One of the most promising platforms for QIP is linear optics \cite{munro05,obrien07,walmsley08,aspelmeyer08,politi08}. For this reason let us now focus on the optimization of the proposed procedure for linear optics. Recently Kieling \it et al.\normalfont\, \cite{kieling10} have identified the maximum success probability of a c-phase in the framework of linear optics as
\begin{eqnarray}
   P_{C}(\varphi) =
   \left(1+2\left|\sin\frac{\varphi}{2}\right|+2^{3/2}\sin\frac{\pi-\varphi}{4}{\left|\sin\frac{\varphi}{2}\right|^{1/2}}\right)^{-2},
\label{eq:cphase_psucc}
\end{eqnarray}
which does not depend on the input state. The optimization of the proposed scheme seeks to maximize the success probability of the c-phase gate used for KLM state preparation. With respect to that a numerical simulation (or optimization) has been carried out to reveal the maximum achievable success probability for several KLM states. The target KLM state of presented numerical simulation is the mono-parametric class of two-qubit KLM state motivated by Franson's~\it et al.\normalfont\, definition \cite{franson02} (triangular-shaped amplitude function)
\begin{equation}
\label{eq:klm_franson}
|\psi\rangle_{\subtxt{KLM}} = \alpha_0 |\qnula\qnula\rangle + \alpha_1 |\qjedna\qnula\rangle + \alpha_0 |\qjedna\qjedna\rangle.
\end{equation}

\begin{figure}
\includegraphics[scale=1.1]{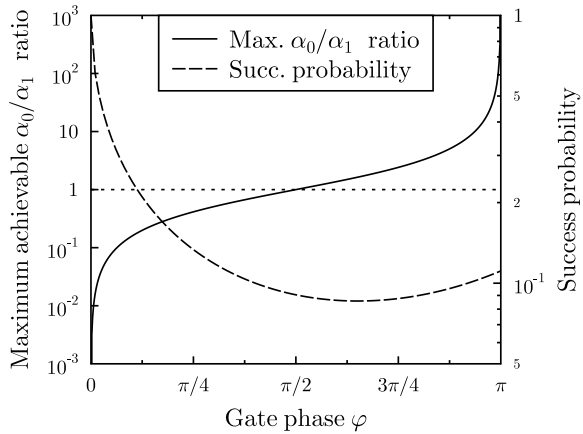}
\caption{Maximum achievable $|\alpha_0/\alpha_1|$ ratio for a given phase shift of the c-phase gate. The success probability of the optimal linear optical c-phase gate as a function of its phase shift is also depicted for reference.}
\label{fig:maxGBratio}
\end{figure}

The amplitudes $\alpha_0$ and $\alpha_1$ are now considered to be real numbers as it has been shown above that the phase can always be set by the choice of $\phi_c$ and $\phi_s$. These phases are independent of the gate phase shift $\varphi$ and therefore have no effect on the success probability. The presented optimization will focus on the amplitude ratio $|\alpha_0/\alpha_1|$ and investigate the corresponding success probability. First numerical simulation has been performed to determine the maximum achievable $|\alpha_0/\alpha_1|$ ratio for a given phase shift. Results of this simulation are presented in figure \ref{fig:maxGBratio}. One can observe that maximum achievable $|\alpha_0/\alpha_1|$ ratio grows monotonously with the phase shift $\varphi$. For reference the success probability (\ref{eq:cphase_psucc}) as a function of the phase shift $\varphi$ is also depicted along with the reference ratio $|\alpha_0/\alpha_1| = 1$ corresponding to the original KLM state definition.

\begin{figure}
\includegraphics[scale=1.1]{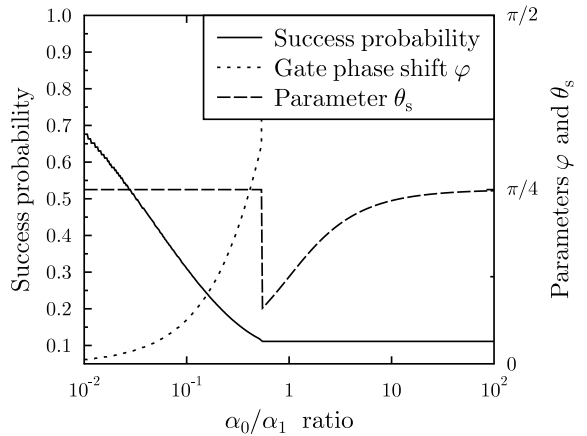}
\caption{Maximum achievable success probability and corresponding optimal $\theta_s$ and $\varphi$ parameters are plotted as a function of $|\alpha_0/\alpha_1|$ ratio. Please note that the optimal setting of $\varphi$ for $|\alpha_0/\alpha_1| > 0.54$ is $\varphi = \pi$ (this explains the step of $\varphi$ at $|\alpha_0/\alpha_1| = 0.54$).}
\label{fig:GBratioPsucc}
\end{figure}

The second numerical simulation has been carried out to determine the maximum achievable success probability for a given $|\alpha_0/\alpha_1|$ ratio (see figure \ref{fig:GBratioPsucc}). Also the setting of the phase shift $\varphi$ and the parameter of the signal qubit $\theta_s$ are depicted to illustrate the optimal strategy. This strategy is different in two regions separated by the amplitude ratio $|\alpha_0/\alpha_1| \approx 0.54$. In the first region ($|\alpha_0/\alpha_1| \leq 0.54$) setting $\theta_s = \frac{\pi}{4}$ and the phase shift $\varphi$ accordingly is the optimal way. One tries to minimize the phase shift used for the KLM state preparation, because the success probability is a decreasing function of the phase shift. To keep the phase shift minimal, one has to set $\theta_s = \frac{\pi}{4}$, because for a given phase shift $\varphi$ the setting $\theta_s = \frac{\pi}{4}$ maximizes the $|\alpha_0/\alpha_1|$ ratio.

On the other hand, in the second region ($|\alpha_0/\alpha_1| > 0.54$) the previously mentioned strategy will not yield optimal results. This is because of the success probability not being monotonous in this region. Setting $\varphi = \pi$ and adjusting the $\theta_s$ instead is the optimal way here.

Both this and the original Franson~\it et al.\normalfont\, scheme requires $n-1$ times using the c-phase gate in order to generate $n$-qubit KLM state. This leads to the overall success probability for $n$-qubit KLM state
\begin{equation}
P_\mathrm{KLM} = \prod_{i=1}^{n-1} P_C(\varphi_i),
\end{equation}
where $n$ denotes the number of qubits and $P_C(\varphi_i)$ is the success probability of the controlled phase gate set for the phase shift $\varphi_i$ used in the $i^{\mathrm{th}}$ repetition of the c-phase gate. The Franson~\it et al.\normalfont\, proposal considers only $\varphi_i = \pi$ for all values of $i$. So for example in the 2-qubit case, the success probability of Franson scheme would yield a constant value of $0.11$ (based on the optimal linear optical controlled phase gate). To emphasize the improvement achieved by the tunability of the phase gate, let us consider an example of $|\alpha_0/\alpha_1| = 0.25$. For this particular choice the success probability of the scheme proposed in this paper would be $0.18$, which is a 60\% improvement. This improvement in success probability varies with the particular choice of the target KLM state (see figure \ref{fig:GBratioPsucc}).
\section{Generalization to $n$-qubit KLM states}
\begin{figure}
\includegraphics[scale=1.1]{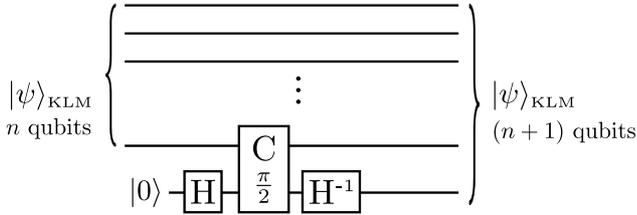}
\caption{Generalization of the two-qubit scheme to an arbitrary number of qubits. Input $n$-qubit KLM state is combined with a new qubit initially in $|\qnula\rangle$ state. $H$ denotes the Hadamard gate and $C$ denotes the c-phase gate (this time set to impose the phase shift $\varphi = \frac{\pi}{2}$).}
\label{fig:scheme_multi}
\end{figure}

The proposed two-qubit scheme can be generalized to prepare KLM states of an arbitrary number of qubits. For simplicity let us now presume all complex amplitudes of the $n$-qubit KLM state being equal (original KLM state definition). To illustrate the generalization procedure the step from two-qubit to three-qubit KLM state is explained and also illustrated in figure \ref{fig:scheme_multi}). Going from two to three qubit KLM state means to perform the following transformation
\begin{eqnarray}
\label{eq:2to3}
|\qnula_1\qnula_2\rangle & \rightarrow & |\qnula_1\qnula_2\qnula_3\rangle \nonumber \\
|\qjedna_1\qnula_2\rangle & \rightarrow & |\qjedna_1\qnula_2\qnula_3\rangle \nonumber \\
|\qjedna_1\qjedna_2\rangle & \rightarrow & |\qjedna_1\qjedna_2\qnula_3\rangle + |\qjedna_1\qjedna_2\qjedna_3\rangle,
\end{eqnarray}
where indexes 1 and 2 denote the first and second original qubits of the two-qubit KLM state and the index 3 denotes the newly added qubit. This transformation can be implemented by addition of a new qubit initially in the state $|\qnula\rangle$. This new qubit is firstly subjected to the Hadamard gate
\begin{equation}
\label{eq:hadamard}
|\qnula\rangle \rightarrow |\qnula\rangle + |\qjedna\rangle.
\end{equation}
After that it is propagated through the c-phase gate set to phase $\varphi = \frac{\pi}{2}$ along with the last of the original KLM qubits. At the end an inverse Hadamard gate is placed in the new qubit mode. One can see that in the case of the last original qubit being $|\qnula\rangle$, the phase shift imposed to the new qubit is zero and the new qubit leaves the scheme in the state $|\qnula\rangle$. On the other hand if the last original qubit is in the state $|\qjedna\rangle$ the new qubit gets a $\frac{\pi}{2}$ phase shift and yields $|\qnula\rangle + |\qjedna\rangle$ after leaving the inverse Hadamard gate.

The generalization to an arbitrary number of qubits is straightforward. To generate an $(n+1)$-qubit KLM state from an $n$-qubit KLM state ($n \leq 2$) a new qubit is added at the end of the original qubits and subjected to the procedure described in previous paragraph. The general scheme is depicted in figure \ref{fig:scheme_multi}. 

\section{Optimization of the generalized scheme}
The previous section is just a proof of the scalability of the scheme, but does not give optimal setting with respect to the success probability. A similar optimization as for the two-qubit KLM states can be considered to maximize the yield of the scheme. Hadamard gates can be replaced by more general single qubit transformations and together with the tunability of the phase shift imposed by every controlled phase gate the overall success probability can be optimized with respect to the selected target KLM state.

One can use the iterative procedure starting from $n$-qubit KLM state with amplitudes $\alpha_j^\mathrm{[n]}$, $j = 0...n$ and going to $(n+1)$-qubit KLM state with amplitudes $\alpha_j^\mathrm{[n+1]}$, $j = 0...n+1$. Here the upper index denotes the $n$-qubit starting KLM state and $(n+1)$-qubit target KLM state. Note that in this case the c-phase gate is applied to the last of the original qubits ($n^\mathrm{th}$ qubit) and a newly added ($n+1)^\mathrm{th}$ qubit. This new qubit can be expressed in the form of~ $|\psi_s\rangle$ as defined by (\ref{eq:cphase_input}) and the last original qubit takes effectively the form similar to~ $|\psi_c\rangle$ with

\begin{eqnarray}
\cos\theta_c &  = & \sqrt{\sum_{j=0}^{n-1}|\alpha_j^\mathrm{[n]}|^2} \quad \mbox{(corresponding to the $|0\rangle$ state)}\nonumber\\
\sin\theta_c & = & |\alpha_{n}^\mathrm{[n]}| \quad \mbox{(corresponding to the $|1\rangle$ state)} \nonumber\\
\phi_c & = & \arg\left(\alpha_{n}^\mathrm{[n]}\right)
\label{eq:nmapping}
\end{eqnarray}
also following the original definition (\ref{eq:cphase_input}). With this mapping one can proceed in the similar way as explicitly described in the second section. The resulting amplitudes~ $\alpha_j^\mathrm{[n+1]}$ are then in the form
\begin{eqnarray}
\alpha_j^\mathrm{[n+1]} & = & \alpha_j^\mathrm{[n]}, \mathrm{for\,} j = 0...n-1 \nonumber\\
\alpha_{n}^\mathrm{[n+1]} & = & |\alpha_{n}^\mathrm{[n]}|\mathrm{e}^{i\phi_c} \tau \nonumber\\
\alpha_{n+1}^\mathrm{[n+1]} & = & |\alpha_{n}^\mathrm{[n]}|\mathrm{e}^{i\phi_c} \epsilon,
\end{eqnarray}
where $\phi_c$ is defined by (\ref{eq:nmapping}) and $\tau$ and $\epsilon$ by (\ref{eq:tau_epsilon}). The equations become increasingly complicated with the growing number of qubits. For this reason one can seek the solution numerically.

As a result of such a numerical optimization, one can for example prepare a 4-qubit KLM state of the triangular-shaped amplitudes in the form of 
\begin{eqnarray}
|\psi\rangle_{\subtxt{KLM}} = \frac{1}{N}\sum_{j=0}^4 \alpha_j |\qjedna\rangle^j |\qnula\rangle^{n-j} \\
\alpha_0 = \alpha_4 = 1, \alpha_1 = \alpha_3 = 3, \alpha_2 = 6
\end{eqnarray}
($N = \sqrt{\sum_{j=0}^n |\alpha_j|^2}$) with the success probability of 0.19\% while the original proposal would give only 0.14\% success probability (40\% improvement). Note that this improved success probability would allow almost 1.5 times higher rate of preparation of KLM states for the "nearly deterministic" protocol proposed by Knill, Laflamme and Milburn \cite{knill01}. The reason for the improvement in the success probability is the fact that using a tunable phase shift, one can operate the controlled phase gate at optimal phase shift. Because one can always set the gate to operate at the phase $\pi$ and set single qubit operations accordingly, the proposed scheme would never give lower success probability as the one proposed by Franson~ \it et al\normalfont.\, The optimal strategy for setting the phase shift imposed by the gate in every step of the generalized procedure is similar to the strategy discussed in the Sec. III for the 2-qubit case. This can be summarized by an inequality
\begin{eqnarray}
P_\mathrm{KLM_Franson} = \prod_{i=1}^{n-1} P_C(\pi) = \nonumber \\
= P_C(\pi)^{n-1} \leq P_\mathrm{KLMnew} = \prod_{i=1}^{n-1} P_C(\varphi_i),
\end{eqnarray}
where the left-hand side corresponds to the success probability of the Franson~ \it et al.\normalfont\, proposal and the right-hand side corresponds to the success probability of the scheme described in this paper. In the worst case scenario hereby proposed scheme allows to set $\varphi = \pi$ to generate any KLM state and in this case the inequality would be saturated.

\section{Conclusions}
The scheme presented in this paper shows how a tunable controlled phase gate can be used to generate arbitrary $n$-qubit KLM states. In comparison with the Franson~ \it et al.\normalfont\, proposal, this scheme gives higher success probability depending of the requested KLM state. It can offer a significant improvement in generation of ancillary states for efficient quantum computing. Please note that this paper discusses the improved generation success probability (rate) for the KLM ancillary states. It should not be confused with the success probability of the teleportation based KLM scheme that employs these ancillary states and considers them as already prepared. Several specific KLM states are discussed in this paper and their preparation success probabilities shown to demonstrate this improvement.

\section{Acknowledgement}
The author would like to thank his colleagues Jan Soubusta and Anton\'in \v Cernoch for fruitful discussion on the subject of this paper.

The author gratefully acknowledge the support by the Operational Program Research and
Development for Innovations - European Regional Development Fund (project CZ.1.05/2.1.00/03.0058  and the  Operational Program Education for Competitiveness - European Social Fund (project CZ.1.07/2.3.00/20.0017 of the Ministry of Education, Youth and Sports of the Czech Republic, by Palacky University (internal grant PrF-2011-009).

This paper is dedicated to my girlfriend Barborka.
\\

\end{document}